\RequirePackage{lineno}
\documentclass[aps,prd,twocolumn,superscriptaddress]{revtex4}

\usepackage{graphicx}

\newcommand{\met}{$E_T\hspace{-1.1em}/$\hspace{0.7em}}

\begin{document}

%\preprint{}

\title{Measurement of event shapes in $p\bar p$ collisions at $\sqrt{s}=1.96$ TeV}

%\author{The CDF collaboration}
%\thanks{}
\affiliation{}


\date{\today}

\begin{abstract}
A study of event shape observables in proton-antiproton collisions at $\sqrt{s}$=1.96 TeV is presented. The data for this analysis were recorded by the CDF II detector at the Tevatron collider. The variables studied are the transverse thrust and thrust minor, both defined in the plane perpendicular to the beam direction. The observables are measured using energies from unclustered calorimeter cells. In addition to studies of the differential distributions, we present the dependence of event shape mean values on the leading jet transverse energy.  Data are compared with {\sc pythia} Tune A and to resummed parton level predictions that were matched to fixed order results at NLO accuracy (NLO+NLL). Predictions from {\sc pythia} Tune A agree fairly well with the data. However, the underlying event contributes significantly to these observables, making it difficult to make direct comparisons to the NLO+NLL predictions, which do not account for the underlying event. To overcome this difficulty, we introduce a new observable, a weighted difference of the mean values of the thrust and thrust minor, which is less sensitive to the underlying event, allowing for a comparison with NLO+NLL. Both {\sc pythia} Tune A and the NLO+NLL calculations agree well within the 20$\%$ theoretical uncertainty with the data for this observable, indicating that perturbative QCD successfully describes shapes of the hadronic final states.
\end{abstract}
%\pacs{}
%\keywords{}
%\interfootnotelinepenalty=10000
\raggedbottom 
\affiliation{Institute of Physics, Academia Sinica, Taipei, Taiwan 11529, Republic of China} 
\affiliation{Argonne National Laboratory, Argonne, Illinois 60439, USA} 
\affiliation{University of Athens, 157 71 Athens, Greece} 
\affiliation{Institut de Fisica d'Altes Energies, Universitat Autonoma de Barcelona, E-08193, Bellaterra (Barcelona), Spain} 
\affiliation{Baylor University, Waco, Texas 76798, USA} 
\affiliation{Istituto Nazionale di Fisica Nucleare Bologna, $^z$University of Bologna, I-40127 Bologna, Italy} 
\affiliation{University of California, Davis, Davis, California 95616, USA} 
\affiliation{University of California, Los Angeles, Los Angeles, California 90024, USA} 
\affiliation{Instituto de Fisica de Cantabria, CSIC-University of Cantabria, 39005 Santander, Spain} 
\affiliation{Carnegie Mellon University, Pittsburgh, Pennsylvania 15213, USA} 
\affiliation{Enrico Fermi Institute, University of Chicago, Chicago, Illinois 60637, USA}
\affiliation{Comenius University, 842 48 Bratislava, Slovakia; Institute of Experimental Physics, 040 01 Kosice, Slovakia} 
\affiliation{Joint Institute for Nuclear Research, RU-141980 Dubna, Russia} 
\affiliation{Duke University, Durham, North Carolina 27708, USA} 
\affiliation{Fermi National Accelerator Laboratory, Batavia, Illinois 60510, USA} 
\affiliation{University of Florida, Gainesville, Florida 32611, USA} 
\affiliation{Laboratori Nazionali di Frascati, Istituto Nazionale di Fisica Nucleare, I-00044 Frascati, Italy} 
\affiliation{University of Geneva, CH-1211 Geneva 4, Switzerland} 
\affiliation{Glasgow University, Glasgow G12 8QQ, United Kingdom} 
\affiliation{Harvard University, Cambridge, Massachusetts 02138, USA} 
\affiliation{Division of High Energy Physics, Department of Physics, University of Helsinki and Helsinki Institute of Physics, FIN-00014, Helsinki, Finland} 
\affiliation{University of Illinois, Urbana, Illinois 61801, USA} 
\affiliation{The Johns Hopkins University, Baltimore, Maryland 21218, USA} 
\affiliation{Institut f\"{u}r Experimentelle Kernphysik, Karlsruhe Institute of Technology, D-76131 Karlsruhe, Germany} 
\affiliation{Center for High Energy Physics: Kyungpook National University, Daegu 702-701, Korea; Seoul National University, Seoul 151-742, Korea; Sungkyunkwan University, Suwon 440-746, Korea; Korea Institute of Science and Technology Information, Daejeon 305-806, Korea; Chonnam National University, Gwangju 500-757, Korea; Chonbuk National University, Jeonju 561-756, Korea} 
\affiliation{Ernest Orlando Lawrence Berkeley National Laboratory, Berkeley, California 94720, USA} 
\affiliation{University of Liverpool, Liverpool L69 7ZE, United Kingdom} 
\affiliation{University College London, London WC1E 6BT, United Kingdom} 
\affiliation{Centro de Investigaciones Energeticas Medioambientales y Tecnologicas, E-28040 Madrid, Spain} 
\affiliation{Massachusetts Institute of Technology, Cambridge, Massachusetts 02139, USA} 
\affiliation{Institute of Particle Physics: McGill University, Montr\'{e}al, Qu\'{e}bec, Canada H3A~2T8; Simon Fraser University, Burnaby, British Columbia, Canada V5A~1S6; University of Toronto, Toronto, Ontario, Canada M5S~1A7; and TRIUMF, Vancouver, British Columbia, Canada V6T~2A3} 
\affiliation{University of Michigan, Ann Arbor, Michigan 48109, USA} 
\affiliation{Michigan State University, East Lansing, Michigan 48824, USA}
\affiliation{Institution for Theoretical and Experimental Physics, ITEP, Moscow 117259, Russia}
\affiliation{University of New Mexico, Albuquerque, New Mexico 87131, USA} 
\affiliation{Northwestern University, Evanston, Illinois 60208, USA} 
\affiliation{The Ohio State University, Columbus, Ohio 43210, USA} 
\affiliation{Okayama University, Okayama 700-8530, Japan} 
\affiliation{Osaka City University, Osaka 588, Japan} 
\affiliation{University of Oxford, Oxford OX1 3RH, United Kingdom} 
\affiliation{Istituto Nazionale di Fisica Nucleare, Sezione di Padova-Trento, $^{aa}$University of Padova, I-35131 Padova, Italy} 
\affiliation{LPNHE, Universite Pierre et Marie Curie/IN2P3-CNRS, UMR7585, Paris, F-75252 France} 
\affiliation{University of Pennsylvania, Philadelphia, Pennsylvania 19104, USA}
\affiliation{Istituto Nazionale di Fisica Nucleare Pisa, $^{bb}$University of Pisa, $^{cc}$University of Siena and $^{dd}$Scuola Normale Superiore, I-56127 Pisa, Italy} 
\affiliation{University of Pittsburgh, Pittsburgh, Pennsylvania 15260, USA} 
\affiliation{Purdue University, West Lafayette, Indiana 47907, USA} 
\affiliation{University of Rochester, Rochester, New York 14627, USA} 
\affiliation{The Rockefeller University, New York, New York 10065, USA} 
\affiliation{Istituto Nazionale di Fisica Nucleare, Sezione di Roma 1, $^{ee}$Sapienza Universit\`{a} di Roma, I-00185 Roma, Italy} 

\affiliation{Rutgers University, Piscataway, New Jersey 08855, USA} 
\affiliation{Texas A\&M University, College Station, Texas 77843, USA} 
\affiliation{Istituto Nazionale di Fisica Nucleare Trieste/Udine, I-34100 Trieste, $^{ff}$University of Trieste/Udine, I-33100 Udine, Italy} 
\affiliation{University of Tsukuba, Tsukuba, Ibaraki 305, Japan} 
\affiliation{Tufts University, Medford, Massachusetts 02155, USA} 
\affiliation{University of Virginia, Charlottesville, VA  22906, USA}
\affiliation{Waseda University, Tokyo 169, Japan} 
\affiliation{Wayne State University, Detroit, Michigan 48201, USA} 
\affiliation{University of Wisconsin, Madison, Wisconsin 53706, USA} 
\affiliation{Yale University, New Haven, Connecticut 06520, USA} 
\author{T.~Aaltonen}
\affiliation{Division of High Energy Physics, Department of Physics, University of Helsinki and Helsinki Institute of Physics, FIN-00014, Helsinki, Finland}
\author{B.~\'{A}lvarez~Gonz\'{a}lez$^v$}
\affiliation{Instituto de Fisica de Cantabria, CSIC-University of Cantabria, 39005 Santander, Spain}
\author{S.~Amerio}
\affiliation{Istituto Nazionale di Fisica Nucleare, Sezione di Padova-Trento, $^{aa}$University of Padova, I-35131 Padova, Italy} 

\author{D.~Amidei}
\affiliation{University of Michigan, Ann Arbor, Michigan 48109, USA}
\author{A.~Anastassov}
\affiliation{Northwestern University, Evanston, Illinois 60208, USA}
\author{A.~Annovi}
\affiliation{Laboratori Nazionali di Frascati, Istituto Nazionale di Fisica Nucleare, I-00044 Frascati, Italy}
\author{J.~Antos}
\affiliation{Comenius University, 842 48 Bratislava, Slovakia; Institute of Experimental Physics, 040 01 Kosice, Slovakia}
\author{G.~Apollinari}
\affiliation{Fermi National Accelerator Laboratory, Batavia, Illinois 60510, USA}
\author{J.A.~Appel}
\affiliation{Fermi National Accelerator Laboratory, Batavia, Illinois 60510, USA}
\author{A.~Apresyan}
\affiliation{Purdue University, West Lafayette, Indiana 47907, USA}
% ...
\collaboration{CDF Collaboration\footnote{ With visitors from $^a$University of Massachusetts Amherst, Amherst, Massachusetts 01003,
$^b$Istituto Nazionale di Fisica Nucleare, Sezione di Cagliari, 09042 Monserrato (Cagliari), Italy,
$^c$University of California Irvine, Irvine, CA  92697, 
$^d$University of California Santa Barbara, Santa Barbara, CA 93106
$^e$University of California Santa Cruz, Santa Cruz, CA  95064,
$^f$CERN,CH-1211 Geneva, Switzerland,
$^g$Cornell University, Ithaca, NY  14853, 
$^h$University of Cyprus, Nicosia CY-1678, Cyprus, 
$^i$University College Dublin, Dublin 4, Ireland,
$^j$University of Fukui, Fukui City, Fukui Prefecture, Japan 910-0017,
$^k$Universidad Iberoamericana, Mexico D.F., Mexico,
$^l$Iowa State University, Ames, IA  50011,
$^m$University of Iowa, Iowa City, IA  52242,
$^n$Kinki University, Higashi-Osaka City, Japan 577-8502,
$^o$Kansas State University, Manhattan, KS 66506,
$^p$University of Manchester, Manchester M13 9PL, England,
$^q$Queen Mary, University of London, London, E1 4NS, England,
$^r$Muons, Inc., Batavia, IL 60510,
$^s$Nagasaki Institute of Applied Science, Nagasaki, Japan, 
$^t$National Research Nuclear University, Moscow, Russia,
$^u$University of Notre Dame, Notre Dame, IN 46556,
$^v$Universidad de Oviedo, E-33007 Oviedo, Spain, 
$^w$Texas Tech University, Lubbock, TX  79609, 
$^x$Universidad Tecnica Federico Santa Maria, 110v Valparaiso, Chile,
$^y$Yarmouk University, Irbid 211-63, Jordan,
$^{gg}$On leave from J.~Stefan Institute, Ljubljana, Slovenia, 
}}
\noaffiliation

\maketitle
\section{Introduction}
The hadronic final states produced in hard collisions can be characterized by a number of variables that describe the distribution of outgoing particles in the event. These are referred to as event-shape variables. Measurements of these variables in $e^{+}e^{-}$ and deep-inelastic scattering experiments \cite{ES_past} allowed a determination of the strong coupling constant $\alpha_{s}$ and its renormalization group running  \cite{alpha_s}, color factor fits of the quantum chromodynamics (QCD) gauge group \cite{CFits}, and more recently, studies of non-perturbative corrections to QCD reviewed in \cite{nPertCorr}.  The goal of the analysis here is to measure event shapes in proton-antiproton collisions and to study their dependence on the transverse energy of the leading jet. The data are compared with {\sc pythia} Tune A \cite{Pythia,PythiaTuneA} and to resummed next-to-leading-logarithm parton level predictions that were matched to fixed order results at next-to-leading-order accuracy \cite{Banfi:2004nk} (referred to as NLO+NLL). This study contributes to our understanding of the underlying event (UE) in a hard-scattering process, and illustrates the need to include underlying event contributions when comparing data with perturbative QCD in hadron-hadron collisions. 

In general, event-shape variables describe geometric properties of the energy flow in QCD final states. They are related to jet-finding algorithms, which are used to categorize events according to their topology. However, they differ significantly in that event-shape variables encode information about the energy flow in the overall event. A single parameter can describe, for example, the transition between a configuration with all particles flowing along a single axis and a configuration where the energy is distributed uniformly over solid angle. Event-shape variables therefore provide an alternative way to characterize an event compared to others based on jet-finding algorithms. Furthermore, they have the advantage of being free of the arbitrariness associated with jet definition, i.e., being either cone or cluster in type, cone sizes, splitting/merging fractions, etc.

The earliest studies of event shapes in hadron-hadron collisions were performed at the ISR \cite{ISR} and the SPS \cite{SPS} in the late 1970s and focused on tracing the emergence of jet-like structures. A decade later event shapes were measured at a hadron-hadron collider during Run I at the Tevatron, where variants of jet broadening and thrust were measured by CDF \cite{CDFold} and D0 \cite{D0old}. Absent from all of these studies was a direct comparison to perturbative QCD. This was in large part due to the intrinsic theoretical difficulties associated with modeling of the hadron collider environment. However, recently a number of tools for investigating event-shape variables beyond leading order have been developed \cite{Banfi:2004nk}, allowing for comparisons with hadron collider data.

In this paper, we report a measurement of event-shape variables using energies from unclustered calorimeter cells in events with transverse energy of the highest-energy (leading) jet ranging from 100 to 300 GeV. Events were produced at the Tevatron collider in proton-antiproton collisions at a center-of-mass energy of 1.96 TeV and were recorded by the CDF II \cite{CDFII} detector. The data used in this analysis were collected from February 2002 to August 2004, with an integrated luminosity of 385 pb$^{-1}$. Section II contains a brief description of the NLO+NLL theoretical predictions used in this analysis. In Sec.~III the observables of interest are introduced and the effects of hadronization and the underlying event are examined. Section~IV outlines the features of the CDF II detector relevant for this analysis. Event pre-selection and the reconstructed objects used in the analysis are described in Sec.~V. Section~VI reviews the measurement, describes instrumental effects on the measurements, and explains how the measurements are corrected for these effects. Section~VII lists the sources of systematic uncertainties that affect the final results.  Finally, results are presented and summarized in Secs.~VIII and IX, respectively.\\

\section{NLO+NLL Theory}

The calculation of event-shape-variable distributions at the parton level in perturbative QCD is divided into two regimes: fixed order and resummed calculations \cite{Banfi:2004nk}. Event shape observables considered in this analysis have the property that large values of the variable coincide with the emission of one or more hard partons at large angles relative to the parent parton. In this regime, the distribution is well described by a traditional perturbative expansion in powers of the strong coupling, $\alpha_{s}$. This method provides an accurate description over most of the range of the variables, e.g. transverse thrust discussed later in this paper.

The method breaks down for small values of the event-shape variables. In fact, all fixed order calculations diverge in the limit that the event shape variable goes to zero. In this region, the differential cross section is primarily sensitive to gluon emissions that are soft compared to the hard scale of the event and/or collinear with one of the hard partons. Such radiation has relatively large emission probabilities due to logarithmic enhancements. In this case, each power of $\alpha_{s}$ in the perturbative expansion is accompanied by a coefficient that grows as $\ln^{2}{1/y}$, where $y$ is the variable of interest. This enhances the importance of higher order terms in the series and the naive requirement that $\alpha_{s}$ be small is no longer sufficient to render these terms negligible. To obtain meaningful predictions in the region $y \to 0$ it is necessary to perform an all-orders resummation of the enhanced logarithmic terms, which can be performed with next-to-leading-logarithm (NLL) precision.

The parton level theoretical predictions used in this analysis correspond to fixed-order results at next-to-leading-order accuracy matched to resummed results at next-to-leading-logarithm accuracy. Theoretical fixed order results are obtained from the Monte-Carlo integration program  {\sc nlojet++} \cite{NLOJET}, while the resummed results arise from the Computer Automated Expert Semi-Analytical Resummer ({\sc CAESAR}) \cite{Banfi:2004yd,Banfi:2004je}. These theoretical calculations include initial and final-state radiation, but do not include multiple parton interactions or beam remnant models. They all use CTEQ 6.1  \cite{CTEQ61} parton distribution functions (PDFs). 

Another technical restriction of NLL calculations is that they apply only to ``global'' variables (i.e., are sensitive to radiation in all directions). This requirement is in direct conflict with the realities of any collider experiment; namely the limited detector coverage at large rapidities. However, the variables considered in this paper are defined exclusively in the transverse plane perpendicular to the beam axis. Therefore, for sufficiently large values of the maximum accessible rapidity, the contribution from the excluded kinematic region is expected to be small and the full global predictions for the studied variables should remain valid for $\ln(y) \le k \eta_\mathit{max}$, where $\eta_\mathit{max}$ is the maximum detector pseudorapidity coverage and $k$ is a constant dependent on the variable $y$~\cite{Banfi:2004nk}.

\section{Definitions of the variables}
\subsection{Transverse Thrust and Thrust Minor}

The transverse-thrust variable, in analogy to the usual thrust, is defined as \cite{ES_past}:
\begin{eqnarray}\label{thrust}
T_{\perp} \equiv \max_{\vec{n}_{T}} \frac {\displaystyle\sum_{i=1}^{n}\left| \vec{q_{\perp,i}} \cdot \vec{n}_{T} \right|} {\displaystyle\sum_{i=1}^{n}\left| \vec{q_{\perp,i}} \right|},
\end{eqnarray}
where $q_{\perp,i}$ is the transverse momentum of the i'th object, where the object is either an outgoing parton, an outgoing stable particle, or a calorimeter cell. The sum runs over all objects in the final state and the thrust axis $\vec{n}_{T}$ is defined as the unit vector in the plane perpendicular to the beam direction that maximizes this expression. For an event with only two back-to-back outgoing objects $T_{\perp} = 1$. In the case of a perfectly cylindrically symmetric event the transverse thrust takes on the value $T_{\perp} = 2/\pi\approx 0.637$. Historically the majority of event-shape variables are defined so that they vanish in the limit of two back-to-back objects. Therefore it is convenient to define $\tau \equiv 1-T_{\perp}$, which has this property. Hereafter, any discussion of the variable called thrust shall refer to the quantity $\tau$, where $0\leq\tau\leq 1-0.637$.
 
The thrust axis $\vec{n}_{T}$ and the beam direction $\hat{z}$ together define the event plane in which the primary hard scattering occurs. The transverse thrust minor is defined as:
\begin{eqnarray}\label{thrustminor}
T_{min} \equiv \frac {\displaystyle\sum_{i=1}^{n}\left| \vec{q_{\perp,i}} \cdot \vec{n}_{m} \right|} {\displaystyle\sum_{i=1}^{n}\left| \vec{q_{\perp,i}} \right|},
\end{eqnarray}
where $\vec{n}_{m}=\vec{n}_{T} \times \hat{z}$. The observable $T_{min}$ is a measure of the out-of-plane transverse momentum and varies from zero for an event entirely in the event plane to $2/\pi$ for a cylindrically symmetric event.

It should be noted that the authors of Ref.~\cite{Banfi:2004nk} also proposed an alternative definition for event-shape variables at hadron colliders to specifically deal with the issue of limited detector coverage. As originally envisioned, the event-shape variables were to be defined over outgoing objects in a reduced central region and rendered ``indirectly'' global by the addition of a ``recoil'' term event-by-event. The ``recoil'' term is defined by momenta in the same central region, but would introduce an indirect sensitivity to momenta outside the region. The proposed recoil term is essentially the vector sum of the transverse momenta in this central region (which by conservation of momentum is equal to the vector sum of transverse momenta outside the region). However, Monte Carlo studies done for this analysis showed that there was almost no correlation between the event-shape variables and the recoil term. As a result, this alternative definition was not pursued.

A number of other event-shape variables (broadenings, hemisphere masses, etc.) whose definitions include longitudinal components of the final state momenta have also been proposed \cite{Banfi:2004nk}. Our studies indicate that these variables are very sensitive to detector resolution in the forward region (see Sec.~IV). Hence, we focus on a study of the observables $\tau$ and $T_{min}$ defined in the transverse plane. By construction these two quantities are infrared and collinear safe.

\subsection{Hadronization and the Underlying Event}
\begin{figure}[htpb]
\includegraphics[width=3.0in]
{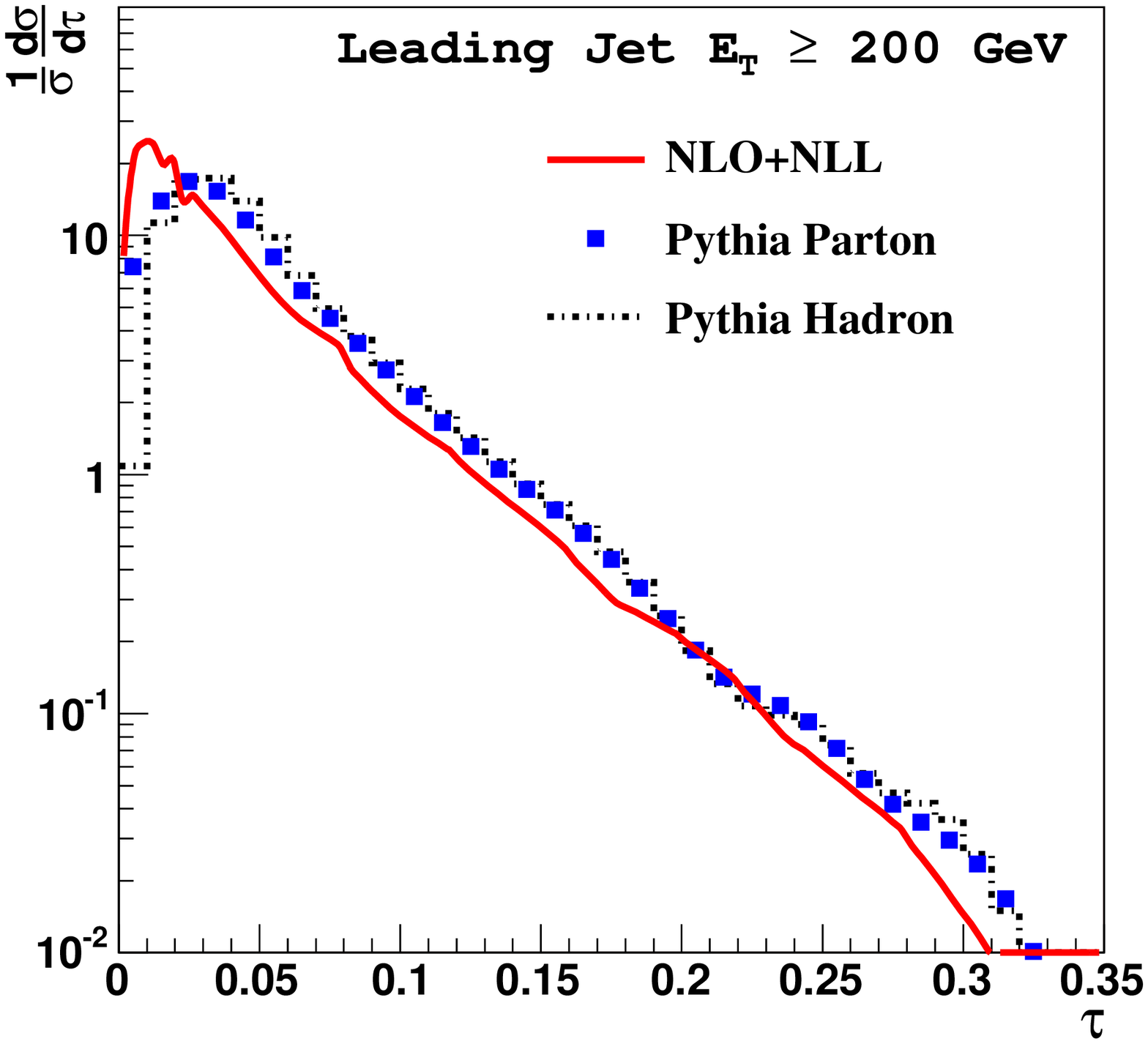}
\includegraphics[width=3.0in]
{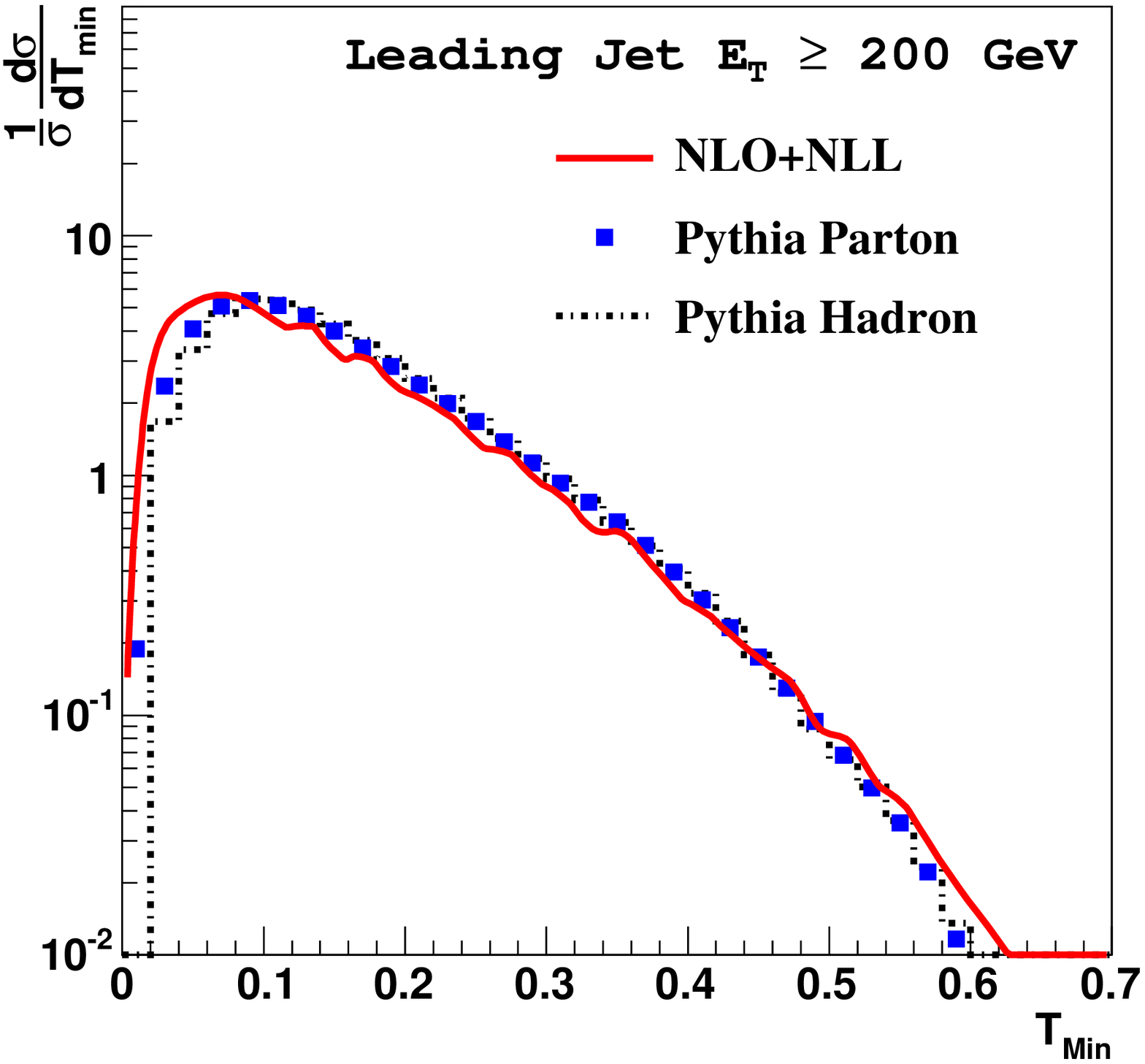}
\caption{Predictions of the transverse thrust and thrust minor distributions for $E_{T}^{lead.jet}$ greater than 200 GeV from a parton level NLO+NLL calculation and from {\sc pythia} without an underlying event at the parton level and without an underlying event at the hadron level (i.e. after hadronization).}
\label{SBZvsPyth}
\end{figure}

The NLO+NLL parton level calculations do not include hadronization effects and they do not have a model of the underlying event.  In particular, they do not include beam-beam remnants nor multiple parton interactions \cite{MPI}.  We use {\sc pythia 6.1} \cite{Pythia} with CTEQ5L PDF sets {\cite{CTEQ5L}} to study the effects of hadronization and the underlying event on the transverse thrust, $\tau$, and thrust minor, $T_{min}$.  The underlying event corresponds to particles that arise from the beam-beam remnants or from multiple-parton interactions. Figure \ref{SBZvsPyth} shows a comparison of the distributions of transverse thrust and thrust minor between the NLO+NLL parton level theoretical predictions, and {\sc pythia} without underlying event at the parton level and after hadronization. The comparison is made for events with leading jet transverse energy $E_{T}^{lead.jet}>200$ GeV; the transverse energy is defined in Sec.~IV. The plot shows that {\sc pythia} and the NLO+NLL parton level predictions have similar shapes for both the transverse thrust and the thrust minor.  However, for the transverse thrust, the {\sc pythia} distribution is shifted toward larger values over the entire range of the variable. Furthermore, hadronization in {\sc pythia} produces only a small shift of the event-shape distributions towards values larger than {\sc pythia} without hadronization, a result expected from LEP \cite{EplusPower}). 

\begin{figure}[htpb]
\includegraphics[width=3.0in]
{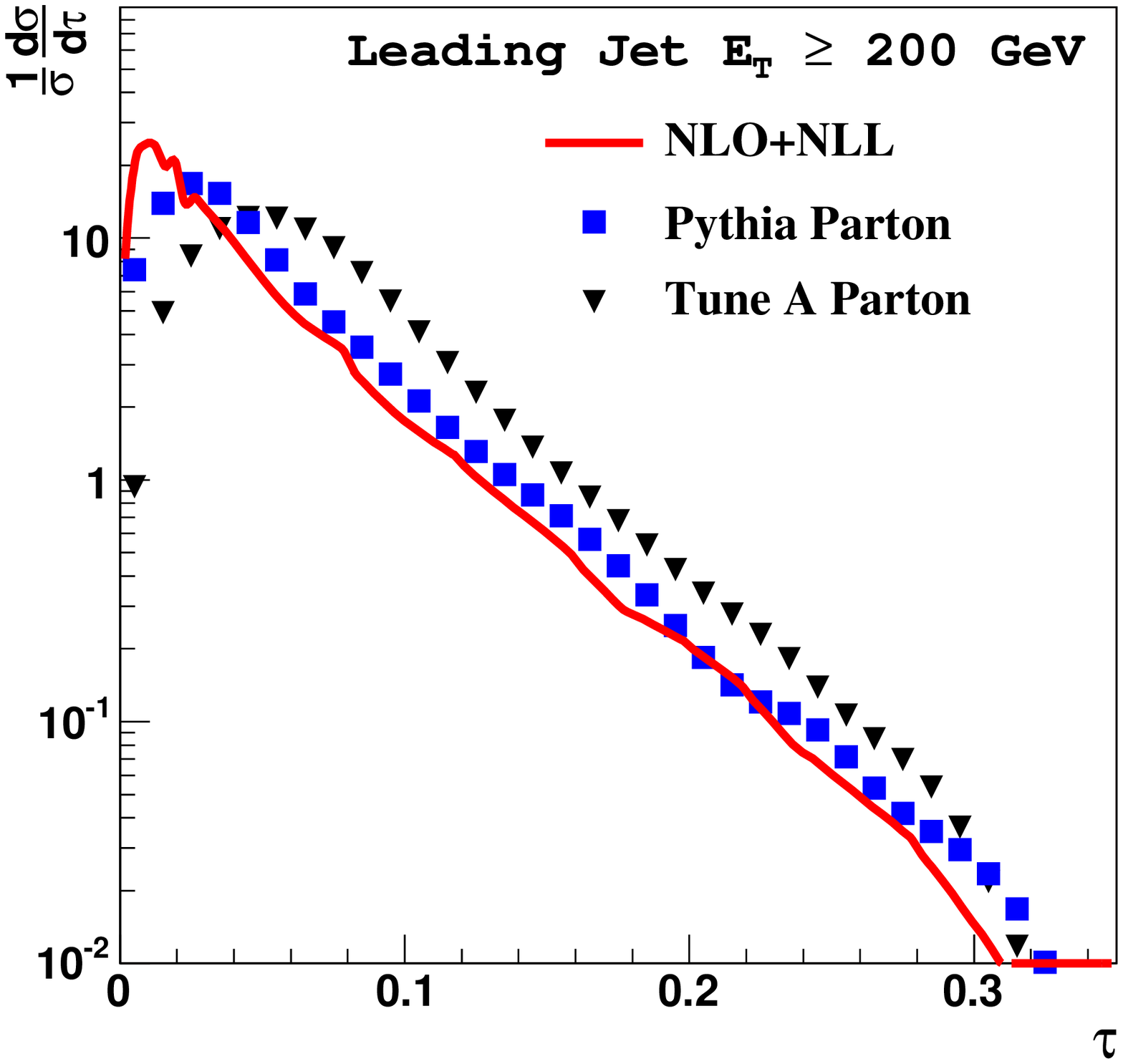}
\includegraphics[width=3.0in]
{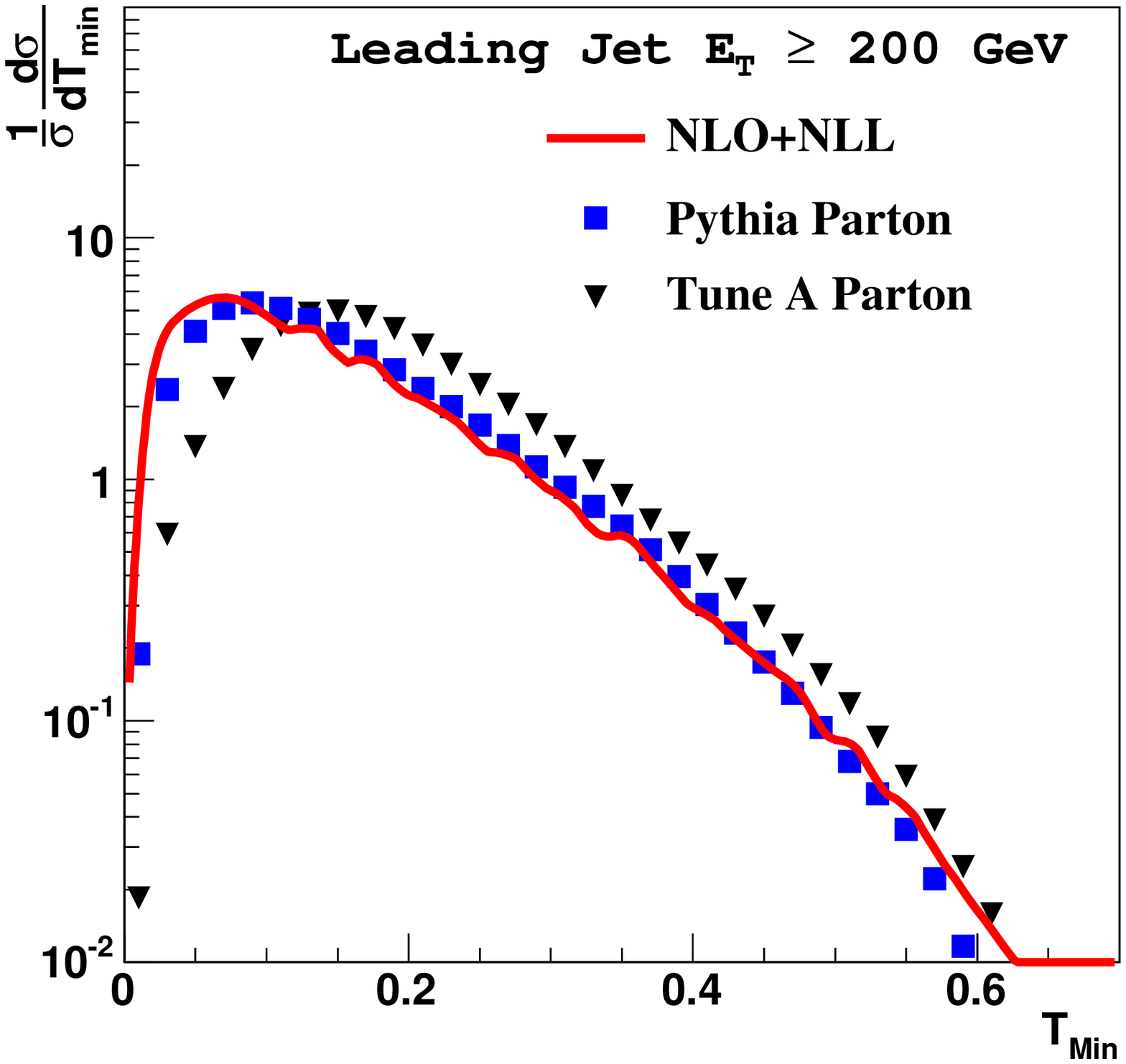}
\caption{Predictions of the transverse thrust and thrust minor distributions for $E_{T}^{lead.jet}$ greater than 200 GeV from a parton-level NLO+NLL calculation and from {\sc pythia} at the parton level without an underlying event and at the parton level with an underlying event (Tune A).}
\label{PythvsTuneA}
\end{figure}

\begin{figure}[htpb]
\includegraphics[width=3.0in]
{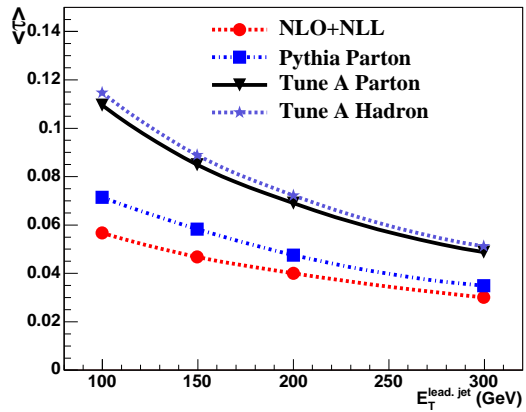}
\includegraphics[width=3.0in]
{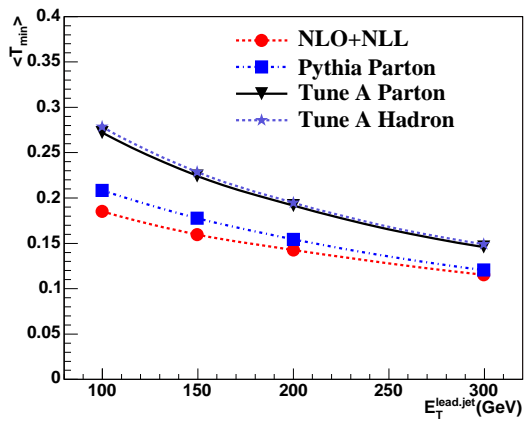}
\caption{Predictions of the mean values of the transverse thrust and thrust minor as a function of the leading jet transverse energy from a parton level NLO+NLL calculation and from {\sc pythia} at the parton level without an UE ({\sc pythia} Parton), at the parton level with an underlying event (Tune A Parton), and at the hadron level with an underlying event (Tune A Hadron).}
\label{PythvsTuneA_LJE}
\end{figure}

Figure \ref{PythvsTuneA} shows a comparison of the event-shape distributions between the NLO+NLL parton level predictions, {\sc pythia} without underlying event, and {\sc pythia} Tune A. {\sc pythia} Tune A includes a model of the underlying event, which was tuned to fit the CDF Run I underlying event data.  We see that the underlying event not only shifts the means towards higher values, but also significantly affects the overall shape of the distributions.  Figure~\ref{PythvsTuneA_LJE} shows mean values as a function of leading jet transverse energy.   There is very little difference in the mean values for {\sc pythia} Tune A at the parton and hadron levels.  The additional partons from multiple-parton interactions saturate the event-shape variable distributions to a point where the ``re-shuffling'' of momenta that occurs at hadronization has little effect on the variable. 

\begin{figure}
\center
\includegraphics[width=3.0in]
{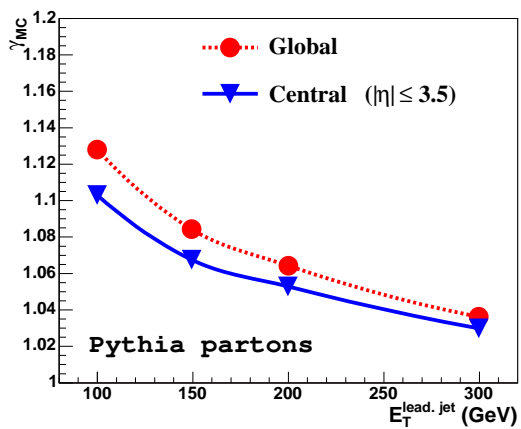}
\caption{Mean value of $\gamma_{MC}$, from Eq.~(6), as obtained from {\sc pythia} at the parton level with and without multiple parton interactions.}
\label{GammaMC}
\end{figure}

\begin{figure}[htpb]
\includegraphics[width=3.0in]
{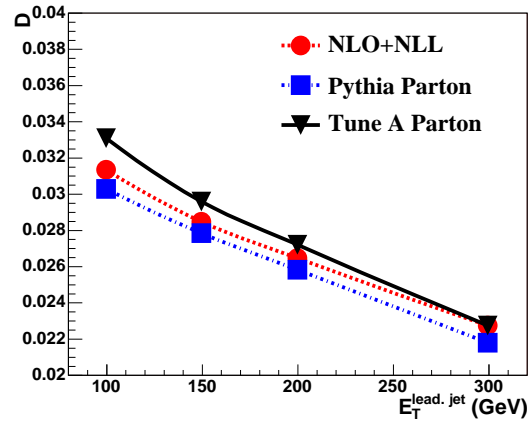}
\caption{Predictions of the mean values of the thrust differential as a function of the leading jet transverse energy from a parton level NLO+NLL calculation and from {\sc pythia} at the parton level with (Tune A) and without an underlying event (Parton). The plot indicates that the underlying event has only a small effect on the mean value of the thrust differential.}
\label{PythvsTuneADiff_LJE}
\end{figure}

We conclude that the underlying event significantly affects the distributions of $\tau$ and $T_{min}$. As a result, a direct comparison of event-shape variable distributions in data with the NLO+NLL parton level predictions is not possible.  However, a quantity less dependent on the underlying event can be constructed from the average values of the thrust and thrust minor.  The dependence of this quantity on the leading jet transverse energy might then allow for a more meaningful comparison between NLO+NLL parton level predictions and the measured data. To this end we begin by considering the definitions of the thrust Eq.~(\ref{thrust}) and thrust minor Eq.~(\ref{thrustminor}). Separating the final state into hard ($q_{\perp}^{hard}$) and soft ($q_{\perp}^{soft}$) components and recognizing that the thrust axis is determined almost entirely by the hard component, we see that the transverse thrust and thrust minor can be written approximately as:
\begin{widetext}
\begin{eqnarray}\label{thrustapprox}
\tau \approx \frac {\displaystyle\sum q_{\perp}^{hard}-\max_{\vec{n_{T}}}\displaystyle\sum q_{\perp}^{hard}\left| \cos\phi^{hard}\right|} {\displaystyle\sum q_{\perp}^{hard} +\displaystyle\sum q_{\perp}^{soft} }+ \frac {\displaystyle\sum q_{\perp}^{soft}( 1 - \left| \cos\phi^{soft}\right|)}  {\displaystyle\sum q_{\perp}^{hard} +\displaystyle\sum q_{\perp}^{soft} },
\end{eqnarray}
\begin{eqnarray}\label{thrustminorapprox}
T_{min} = \frac {\displaystyle\sum q_{\perp}^{hard}\left| \sin\phi^{hard}\right|}  {\displaystyle\sum q_{\perp}^{hard} +\displaystyle\sum q_{\perp}^{soft} }+ \frac {\displaystyle\sum q_{\perp}^{soft}\left| \sin\phi^{soft}\right|}  {\displaystyle\sum q_{\perp}^{hard} +\displaystyle\sum q_{\perp}^{soft} },
\end{eqnarray}
\end{widetext}
where $\phi^{hard}$ and $\phi^{soft}$ represent the angle between the thrust axis and the hard and soft components, respectively. The soft underlying event is expected to be on average uniform over the transverse plane, therefore $1-\tau^{soft}\approx T_{min}^{soft}\approx 2/\pi$. An expression whose numerator is less dependent on the underlying event can be constructed by taking a weighted difference between the mean values of the thrust and thrust minor as follows:
\begin{widetext}
\begin{eqnarray}\label{ourC}
\alpha\left<T_{min}\right>-\beta\left<\tau\right>\approx \alpha \left< \frac {\displaystyle\sum q_{\perp}^{hard}\left| \sin\phi^{hard}\right|}  {\displaystyle\sum q_{\perp}^{hard} +\displaystyle\sum q_{\perp}^{soft} } \right> -\beta\left<\frac {\displaystyle\sum q_{\perp}^{hard}-\max_{\vec{n_{T}}}\displaystyle\sum q_{\perp}^{hard}\left| \cos\phi^{hard}\right|} {\displaystyle\sum q_{\perp}^{hard} +\displaystyle\sum q_{\perp}^{soft}}\right>,
\end{eqnarray}
\end{widetext}
where $\alpha=1-2/\pi$, $\beta=2/\pi$, and the sums run over objects from the hard scattering (``hard'') and from the soft underlying event (``soft''). The soft underlying event in this expression is in the denominator where its contribution is overshadowed by the hard-scattering term. Furthermore, an additional correction factor, $\gamma_{MC}$, can be computed from {\sc pythia} tune A {\cite{Pythia,PythiaTuneA}} generated with and without multiple parton interactions: 
\begin{eqnarray}\label{gamma}
\gamma_{MC}=\frac{\displaystyle\sum \left| q_{\perp}^{NoMPI} \right|+\displaystyle\sum \left| q_{\perp}^{WithMPI} \right| }{\displaystyle\sum \left| q_{\perp}^{NoMPI} \right|}.
\end{eqnarray}
Now we define a new variable:
\begin{eqnarray}\label{D}
D(\left<\tau\right>,\left< T_{min}\right>)=\gamma_{MC}(\alpha\left< T_{min}\right>-\beta\left<\tau\right>),
\end{eqnarray}
which should be less dependent on the underlying event. The $\gamma_{MC}$ correction is applied to the data for comparisons to theory. We call $D(\left<\tau\right>,\left< T_{min}\right>)$ the {\it ``thrust differential''}. It should be insensitive to the underlying event activity in the event, thereby allowing more meaningful comparisons to perturbative calculations of event-shape variables. The variable ranges between 0 and 0.1 and vanishes in both limiting cases of cylindrically symmetric and pencil-like events. The variable allows probing the relative contributions of pQCD and non-pQCD processes to the distributions of the event-shape variables.
It is important to note that the $\gamma_{MC}$ correction factor differs from unity by no more than 13\% over the range of the leading jet $E_{T}^{lead.jet}$ threshold, as shown in Fig.~\ref{GammaMC}. Also, it can be seen in Fig.~\ref{PythvsTuneADiff_LJE} that MC simulations show a strongly-reduced effect of the underlying event on the thrust differential relative to its effect on the transverse thrust and thrust minor shown in Fig.~\ref{PythvsTuneA_LJE}. Experimental measurements of transverse thrust and thrust minor, of the thrust differential, and of the dependence of the latter quantity on the transverse energy threshold of the leading jet are presented in the next sections.

\section{CDF II detector}
Data used in this analysis were recorded with CDF II, a general-purpose detector, designed for precision measurements of the energy, momentum, and trajectories of particles produced in proton-antiproton collisions. This section provides a brief overview of the components relevant to our analysis. A detailed description of the entire detector can be found elsewhere {\cite{CDFII}}.

CDF II uses a spherical coordinate system with the $z$ axis oriented along the proton beam direction and azimuthal angle $\phi$ measured around the beam axis. The polar angle $\theta$  is measured with respect to the positive $z$ (proton-beam) direction used to define pseudorapidity $\eta=-\ln\left[\tan(\frac{\theta}{2})\right]$. 

The CDF II tracking system is used for reconstruction of primary interaction vertices and particle tracks, and is placed inside a 1.4 T solenoidal magnet. An inner, single-sided silicon microstrip detector (Layer~00) is mounted directly on the beam pipe, at an inner radius of $1.15$ cm and an outer radius of $2.1$ cm. A five-layer silicon microstrip detector (SVX II) is situated at the radial distance of 2.5 to 11 cm from the beam line, and consists of three separate barrel modules with a combined length of 96 cm. Three of the five layers combine a $r$-$\phi$ measurement with a $z$-coordinate measurement while the remaining two layers combine $r$-$\phi$ with small-angle stereo views $1.2^{\circ}$. Three additional intermediate silicon layers (ISL) are positioned between 19 and 30 cm from the beam line. The silicon tracker is surrounded by the central outer tracker (COT), an open-cell drift chamber providing up to 96 measurements of a charged particle track over the radial region from 40 to 137 cm. The pseudorapidity region covered by the COT is $\left|\eta\right|<1.0$.

The CDF II tracking system is surrounded by electromagnetic and hadronic calorimeters, whose cells are arranged in a projective tower geometry. The central electromagnetic (CEM), central hadronic (CHA), and wall hadronic calorimeters consist of lead (electromagnetic) and iron (hadronic) layers interspersed with scintillator. The pseudorapidity region covered by these calorimeters is $\left|\eta\right|<1.3$. The segmentation of the central calorimeters is $15^{\circ}$ in $\phi$ and 0.1 units in $\eta$ and again $15^{o}$ in $\phi$ but 0.2 to 0.6 in $\eta$ in a forward ``plug'' region. The measured energy resolutions for the CEM and CHA are $\sigma(E_T)/E_T=13.5\%/\sqrt{E_{T}}\oplus 2\%$ and $\sigma(E_T)/E_T=75\%/\sqrt{E_{T}}\oplus 3\%$, respectively. Here $E_{T}=E\sin{\theta}$ is the transverse energy deposited in a calorimeter tower. Energies are measured in GeV. Additional calorimetry extends the coverage in the forward direction to $\left|\eta\right|<3.6$. The forward electromagnetic calorimeter is constructed of lead and scintillator layers with an energy resolution of $\sigma(E_T)/E_T=16\%/\sqrt{E_{T}}\oplus 1\%$. The forward hadronic calorimeter is made of iron and scintillator layers with an energy resolution of  $\sigma(E_T)/E_T=80\%/\sqrt{E_{T}}\oplus 5\%$.

\section{Event Selection}
\subsection{Triggers}
Events were collected using single-jet triggers with $E_{T}$ thresholds of 50 (J050), 70 (J070), and 100 (J100) GeV. Pre-scale factors are applied to J050 and J070 jet triggers so as not to saturate the available trigger bandwidth; typical values of pre-scale factors are 8 and 50 for J050 and J070, respectively. The J100 trigger is not prescaled. 

\subsection{Jet reconstruction algorithm}
While event shape variables are calculated from unclustered calorimeter cell energies, analyzing their dependence on leading-jet $E_T$ requires use of a jet-energy algorithm for the leading jet scale. Based on calorimeter information, jets are reconstructed using a cone algorithm {\cite{JetClu}}. The algorithm starts with the highest transverse energy tower and forms preclusters from an unbroken chain of continuous seed towers with transverse energies above 1 GeV within a window of $7\times7$ towers centered on the originating seed tower. If a seed tower is outside this window, it is used to form a new precluster. The coordinates of each precluster are the $E_{T}$-weighted sums of $\phi$ and $\eta$ of the seed towers within this precluster. In the next step, all towers with $E_{T}>0.1$ GeV within $R=\sqrt{(\Delta\phi)^{2}+(\Delta\eta)^{2}}=1.0$ of the precluster are merged into a cluster, and its $(\eta,\phi)$-coordinates are recalculated. This procedure of calculating cluster coordinates is iterated until a stable set of clusters is obtained. A cluster is stable when the tower list is unchanged from one iteration to the next. If the clusters have some finite overlap, then an overlap fraction is computed as the sum of the transverse energies of the common towers divided by the $E_T$ of the smaller cluster. If the fraction is above a cutoff value of 0.75, the two clusters are combined. If the fraction is less than the cutoff, the shared towers are assigned to the closer cluster. The raw energy of a jet is the sum of the energies of the towers belonging to the corresponding cluster. Corrections are applied to the raw energy to compensate for the non-linearity and non-uniformity of the response of the calorimeter, the energy deposited inside the jet cone from sources other than the assumed leading jet-parent parton, and the leading parton energy deposited outside the jet cone. A detailed description of this correction procedure can be found in {\cite{JetCluCorr}}. 

\subsection{Offline selection}
Cosmic ray events are rejected by applying a cutoff on the significance of the missing transverse energy \met {\cite{missingEt}}, defined as \met$/\sqrt{\Sigma E_{T}}$, where $\Sigma E_{T}=\Sigma_{i} E_{T}^{i}$ is the total transverse energy of the event, as measured using calorimeter towers with $E_{T}^{i}$ above $100$ MeV. The cutoff values are 5.0, 6.0, and 7.0 GeV$^{1/2}$  for data collected using jet triggers with thresholds of 50, 70, and 100 GeV, respectively.\\ 

To ensure fully efficient vertex and track reconstruction, we require events with a single interaction as evidenced by having only one reconstructed primary interaction vertex with $\left|z\right|<60$ cm. 

Only events with both leading jets in the central region ($\left|\eta\right|<0.7$) are selected. Events are categorized according to the transverse energy threshold of the leading jet as in Ref~\cite{Banfi:2004nk}. The event categories, trigger paths, and number of events after selection criteria are summarized in Table~{\ref{BinSummary}}.

	\begin{table}
	\caption{Summary of the data samples, trigger paths, and number of events present after the offline selection criteria.}
	\begin{ruledtabular}
	\begin{tabular}{ccc}
	$E_{T}^{lead.jet}$ (GeV)  &  Trigger  &   Number of Events    \\
	\\ \hline
	100  &  J050 & ~52546 \\
	150  &  J070 & ~17850 \\
	200  &  J100 & ~26207 \\
	300  &  J100 & ~3126 \\
	\end{tabular}
	\end{ruledtabular}
	\label{BinSummary}
	\end{table}

\section{Measurement and Instrumental Uncertainties}

 The measurement of event-shape variables is performed with unclustered calorimeter towers over the detector's full rapidity range $\left|\eta\right|<3.5$. Towers used in the measurement are required to have minimum E$_{T}$ of 100 MeV. 

\begin{figure}[htpb]
\includegraphics[width=3.0in]
{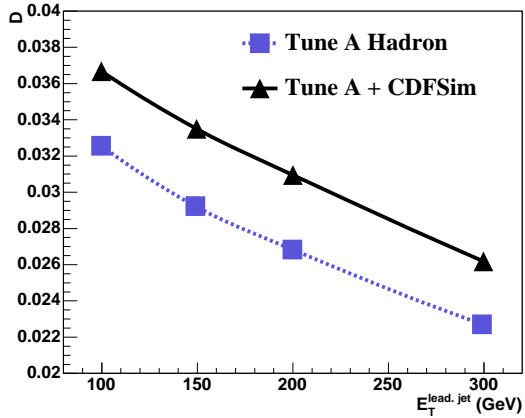}
\caption{The effect of CDF detector simulation on the thrust differential as a function of the leading jet transverse energy.}
\label{DetEffects}
\end{figure}

In general, measurement of the event-shape variables will be distorted by instrumental effects and a correction factor is needed to account for this. Figure \ref{DetEffects} shows the dependence of  $D(\langle \tau \rangle, \langle T_{Min} \rangle)$  on the leading jet transverse energy for {\sc pythia} Tune A at the hadron level and at the calorimeter level after full CDF detector simulation. While the detector effects do not appreciably affect distributions of $\tau$ and $T_{min}$, they induce a significant systematic shift of the thrust differential. Sources for this shift have been studied, and the results are given below.

As a result of the magnetic field, the energy flow of an event as measured by the calorimeter will be broader than in the absence of the field.  To estimate the magnitude of this effect on the thrust differential, MC simulated particles at the hadron level were propagated to the first active layer of the calorimeter under the influence of a 1.4 T  B-field. The direction of each particle was calculated from the $z$ coordinate of the primary interaction vertex and the point of impact on the first calorimeter layer. The effect is found to be $\sim 2\%$ of the values of the thrust differential and negligible compared to the effect of calorimeter granularity described below.
       
To estimate the effect on the thrust differential of the calorimeter energy resolution, we smear the energy of the particles in the MC simulation according to a gaussian with the 1$\sigma$ resolution quoted in section IV. Smearing changes the thrust differential by $< 1\%$ of its value.

Turning now to the effect of the calorimeter granularity, we note that when a particle above threshold is detected, the location returned by the system is the center of the tower and not the exact location of the shower within the tower. As a result, there is an error associated with the granularity of the calorimeter.  In order to understand this effect on the thrust differential, the segmentation of the calorimeter is imposed on MC simulated particles at the hadron level.

\begin{figure}[htpb]
\includegraphics[width=3.0in]
{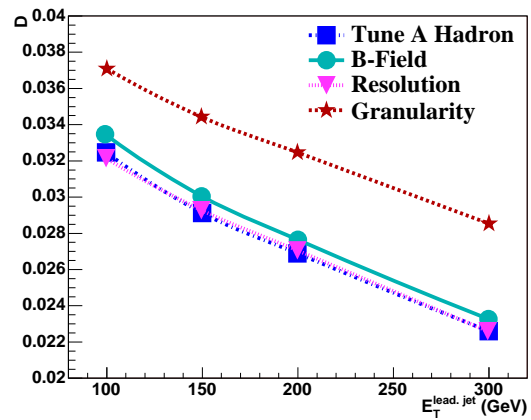}
\caption{Simulation of individual instrumental effects on the thrust differential as a function of the jet energy threshold.}
\label{TauTminDetEffItem}
\end{figure}

The effects of these instrumental errors on the thrust differential are shown in Fig.~\ref{TauTminDetEffItem}. The granularity of the calorimeter is the primary source of the shift of this variable. The shift is taken into account by a bin-by-bin correction to the thrust and thrust minor distributions, which is propagated to $D(\langle \tau \rangle, \langle T_{Min} \rangle)$. 

\begin{figure} [htpb]
\includegraphics[width=3.0in]
{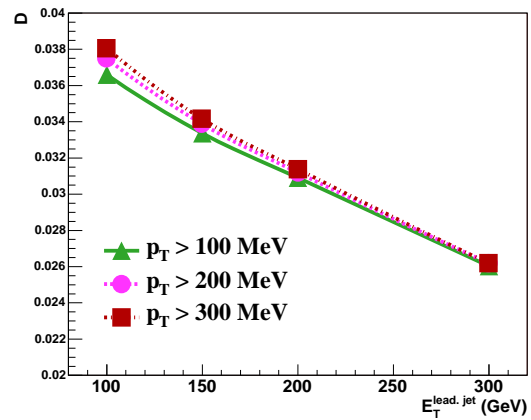}
\caption{Effect of tower E$_{T}$ threshold on the thrust differential as a function of the leading jet transverse energy threshold.}
\label{PTCut}
\end{figure}

In the model calculations referenced here, event-shape variables are defined over all particles in the final state, including those with arbitrarily small momenta.  In order to understand how a cut on the transverse energy affects the variables, we vary the E$_{T}$ threshold on towers from 100 MeV (default) through 200 and 300 MeV. Figure \ref{PTCut} shows that the thrust differential is rather insensitive to the cut on transverse energy at low $q_{\perp}$. While the distributions of thrust and thrust minor get narrower with increasing $E_{T}^{lead.jet}$ threshold, the effect on the thrust differential is negligible compared to the effect of calorimeter granularity.

\section{Systematic Uncertainties}

The sensitivity of the thrust differential to various uncertainties in the event selection procedure is evaluated as follows. For each source of systematic uncertainty, a ``default'' and ``deviated'' variable is constructed. The ``default'' variable is the result of the standard set of cuts defined earlier in this paper, while the ``deviated'' variable is the result of varying a particular parameter within its uncertainty. For each of the data samples corresponding to the four different values of transverse energy thresholds of the leading jet, the systematic uncertainties on the thrust differential are calculated as the difference between the ``deviated'' and the ``default'' results. Each individual source of uncertainty is then added in quadrature to the statistical uncertainty of each data point.

\subsection{Jet Energy Scale}
Theoretical predictions for the event-shape variables are parametrized as functions of the leading jet transverse energy. To evaluate the uncertainty on the leading jet transverse energy due to the jet energy corrections, we use a parametrization that under- and over-estimates the leading jet energy by one standard deviation in the jet energy scale {\cite{JetCluCorr}} and then re-run our event selection.  The difference between the default and the deviated variable is assigned as a systematic uncertainty. As expected, most of the jet energy scale error cancels in the ratio of sums of Eqs.~(1)~and~(2) and in the calculation of D, hence the resulting systematic uncertainty is small.

\subsection{Detector Hermeticity}
The primary interaction vertex is required to lie within 60 cm from the center of the detector in order to ensure that the majority of the event is contained within the detector.  This analysis of event-shape variables uses calorimeter information in the far forward regions of the detector.  As a result, the further a collision occurs from the nominal interaction point the greater the possiblity that particles fall beyond the detector's coverage.  To evaluate the uncertainty due to this effect we require a tighter cut on the $z$ position of the primary vertex, $\left|z\right|<20$ cm.  The difference in the variable between the default and the tight cut is then assigned as a systematic uncertainty.
     
\subsection{Effect of Pile-up}
In the event selection we specifically require events with a single vertex; however, it is possible that two vertices that lie very close to each other are reconstructed as a single vertex.  This ``pile-up'' effect is especially likely at high values of the instantaneous luminosity.  To evaluate the uncertainty due to this effect we separate events in each data sample into high (average of 3 primary interactions per bunch crossing) and low (average of 1.5 primary interactions per bunch crossing) luminosity subsets with approximately equal numbers of events. The thrust differential is then compared between subsets and the difference is taken as a measure of the systematic uncertainty. 

A summary of all uncertainties affecting this measurement are given in Table \ref{FinalPoints}, together with the values of the thrust differential.

\section{Results}

\begin{figure}[htbp]
\includegraphics[width=3.0in]
{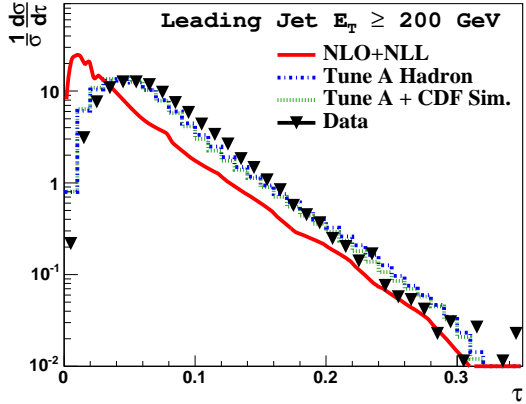}
\includegraphics[width=3.0in]
{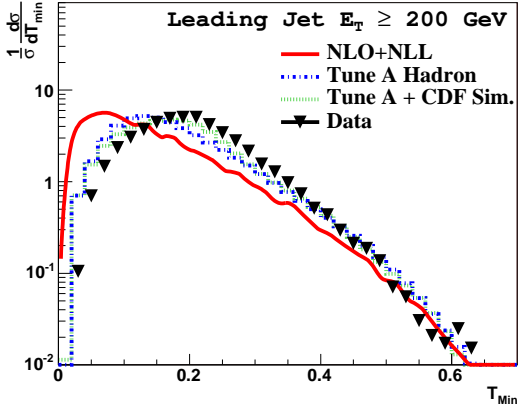}
\caption{ The uncorrected CDF distributions of transverse thrust and thrust minor for leading jet transverse energy greater than 200 GeV. The experimental results are compared with a parton level NLO+NLL calculation and with {\sc pythia} at the hadron level (Tune A Hadron) and at the detector level ({\it i.e.} after CDFSIM).}
\label{TuneAvsData}
\end{figure}

The distributions of the transverse thrust and thrust minor, uncorrected for detector effects, are shown in  Fig.~\ref{TuneAvsData} for events with the $E_{T}^{lead.jet}$ greater than 200 GeV. Distributions for other $E_{T}^{lead.jet}$ thresholds can be found in \cite{PineraThesis}. There is not much difference between {\sc pythia} Tune A at the hadron level and the detector level (CDFSIM). Tune A describes the data fairly well although not perfectly. The distribution of the thrust minor is slightly broader than the Tune A prediction ({\it i.e.} there is slightly more energy out of the plane than predicted by Tune A). The parton level NLO+NLL predictions deviate significantly from the data since they have no underlying event. For events with leading jet transverse energy greater than 200 GeV, the mean value of the $\tau$ distribution shifts from $0.039\pm0.001$ to $0.070\pm0.001$ (parton level NLO+NLL to experiment), while the RMS remains unchanged at $0.040\pm0.001$. The mean value of the $T_{min}$ distribution shifts from $0.142\pm0.002$ to $0.206\pm0.002$ with its RMS decreasing from $0.099\pm0.001$ to $0.087\pm0.001$.

\begin{figure}[htpb]
\includegraphics[width=3.0in]
{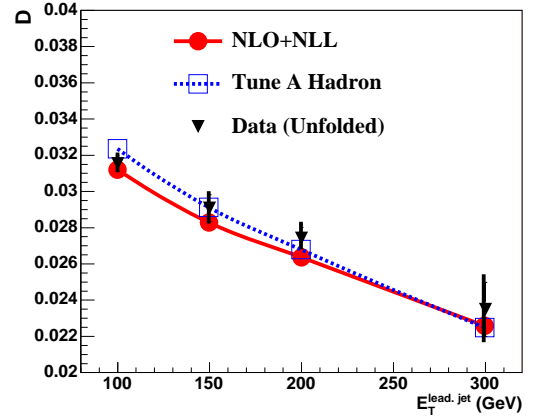}
\caption{The CDF corrected results for the dependence of the thrust differential on the transverse energy of the leading jet. The experimental results are compared with a parton level NLO+NLL calculation and with {\sc pythia} Tune A at the hadron level. The error bars correspond to statistical and systematic uncertainties added in quadrature.}
\label{TuneAvsData_Diff_LJE}
\end{figure}

Figure \ref{TuneAvsData_Diff_LJE} shows the thrust differential as a function of $E_{T}^{lead.jet}$. In this plot the data have been corrected for detector effects. The data are compared with {\sc pythia} Tune A and with parton level NLO+NLL calculations.  The NLO+NLL predictions shown in this figure correspond to a particular choice of renormalization and factorization scale, namely the transverse energy of the leading jet; the theoretical uncertainty on thrust differential is approximately 20\% \cite{Banfi:2010xy,Banfi:PrComm} (theoretical uncertainties on transverse thrust and thrust minor are smaller and are of the order of 10$\%$ or less \cite{Banfi:2010xy}). By construction this observable all but eliminates the sensitivity to the underlying event. Based on $\gamma_{MC}$ variation studies we estimated that the residual effect of the UE is less than few percent. Both {\sc pythia} Tune A and the NLO+NLL calculations agree fairly well with the data, indicating that the non-perturbative effects are small. The corrected data and their associated uncertainties are listed in Table \ref{FinalPoints}.

\begin{table*}[ht]
\caption{Summary of the experimental values of the thrust differential and of its uncertainties.}\label{FinalPoints}
\center
\begin{ruledtabular}
\begin{tabular}{l|c|c|c|c|c}
	$E_{T}^{lead.jet}$ (GeV)  & $D(\langle \tau \rangle, \langle T_{Min} \rangle)$ & Stat. & Jet Energy Scale & Detector Hermeticity & Pile-up \\ \hline
	100 & 315.3$\times 10^{-4}$ &  4.5$\times 10^{-4}$ & 0.3$\times 10^{-4}$ & 0.5$\times 10^{-4}$ & 2.8$\times 10^{-4}$\\
	150 & 290.8$\times 10^{-4}$ &  7.4$\times 10^{-4}$ & 1.3$\times 10^{-4}$ & 1.5$\times 10^{-4}$ & 4.5$\times 10^{-4}$\\
	200 & 275.6$\times 10^{-4}$ &  5.9$\times 10^{-4}$ & 3.4$\times 10^{-4}$ & 6.1$\times 10^{-4}$ & 2.8$\times 10^{-4}$\\
	300 & 235.1$\times 10^{-4}$ &  14.9$\times 10^{-4}$ & 4.7$\times 10^{-4}$ & 7.7$\times 10^{-4}$ & 7.2$\times 10^{-4}$\\
\end{tabular}
\end{ruledtabular}
\end{table*}

\section{Summary}

Event-shape variable distributions are studied using unclustered calorimeter energies in proton-antiproton collisions at a center-of-mass energy of 1.96 TeV. The measurements were performed using individual calorimeter towers with a transverse-energy threshold of 100 MeV. The data are compared to {\sc pythia} Tune A and to resummed parton level predictions that were matched to fixed order results at NLO accuracy (NLO+NLL).  Both the thrust and thrust minor distributions are sensitive to the modeling of the underlying event. The {\sc pythia} Tune A distributions of the observables reproduce the experimental distributions fairly well, although not perfectly.  The data show slightly more energy out of the hard-scattering plane than predicted by Tune A. These observables can be used to improve the modeling of the underlying event.  The NLO+NLL predictions differ significantly from both the data and from  {\sc pythia} Tune A since these calculations do not incorporate either hadronization or the underlying event. 

A new variable, called thrust differential, is introduced. It is a weighted difference of the mean values of the thrust and thrust minor over the event sample. By construction it is less sensitive to the underlying event and hadronization effects. Both {\sc pythia} Tune A and the NLO+NLL calculations succeed in describing the data on the thrust differential. This observable allows a comparison with the NLO+NLL calculations, and data and theory are found to agree well within the 20$\%$ theoretical uncertainty.

\section{Acknowledgements}

	The authors are very grateful to Andrea Banfi, Gavin Salam and Giulia Zanderighi  for collaborative work and for providing us with preliminary results from Ref.~\cite{Banfi:2010xy} prior to its publication. We thank the Fermilab staff and the technical staffs of the participating institutions for their vital contributions. This work was supported by the U.S. Department of Energy and National Science Foundation; the Italian Istituto Nazionale di Fisica Nucleare; the Ministry of Education, Culture, Sports, Science and Technology of Japan; the Natural Sciences and Engineering Research Council of Canada; the National Science Council of the Republic of China; the Swiss National Science Foundation; the A.P. Sloan Foundation; the Bundesministerium f\"ur Bildung und Forschung, Germany; the World Class University Program, the National Research Foundation of Korea; the Science and Technology Facilities Council and the Royal Society, UK; the Institut National de Physique Nucleaire et Physique des Particules/CNRS; the Russian Foundation for Basic Research; the Ministerio de Ciencia e Innovaci\'{o}n, and Programa Consolider-Ingenio 2010, Spain; the Slovak R\&D Agency; and the Academy of Finland.

% Systematic Uncertainties

\begin{references}
\bibitem{missingEt} \met ~is defined as the norm of $-\Sigma_{i} E_{T}^{i} \cdot \vec{n_{i}}$, where $\vec{n_{i}}$ is the unit vector in the azimuthal plane that points from the beamline to the i-th calorimeter tower.
\end{references}
\end{document}